\documentclass[12pt]{iopart}
\usepackage{graphicx}
\usepackage{color}
\begin{document}

\title[ ]{Impact of the protein composition on the structure and viscoelasticity of polymer-like gluten gels}

\author{Laurence Ramos$^{1}$, Am\'elie Banc$^{1}$, Ameur Louhichi$^{1}$, Justine Pincemaille$^{1,2}$, Jacques Jestin$^{3}$, Zhendong Fu$^{4}$, Marie-Sousai Appavou$^{4}$, Paul Menut$^{2,5}$, Marie-H\'el\`ene Morel$^{2}$}

\address{$^{1}$Laboratoire Charles Coulomb (L2C), Univ. Montpellier, CNRS, Montpellier, France}
\address{$^{2}$Ing\'{e}nierie des Agro-polym\`{e}res et Technologies Emergentes (IATE), Univ. Montpellier, CIRAD, INRAE, Montpellier SupAgro, Montpellier, France}
\address{$^{3}$Laboratoire L\'{e}on Brillouin, UMR 12, Université Paris-Saclay, IRAMIS/CEA Saclay, 91191 Gif-sur-Yvette Cedex, France}
\address{$^{4}$Forschungszentrum Jülich GmbH JCNS am MLZ Lichtenbergstr. 1, 85748 Garching, Germany}
\address{$^{5}$Universit\'{e} Paris-Saclay,  INRAE, AgroParisTech, UMR SayFood, 91300 Massy, France}
\ead{laurence.ramos@umontpellier.fr}
\vspace{10pt}
\begin{indented}
\item[]January, 18 2021
\end{indented}

\begin{abstract}
We investigate the structure of gluten polymer-like gels in a binary mixture of water/ethanol, $50/50$ v/v, a good solvent for gluten proteins. Gluten comprises two main families of proteins, monomeric gliadins and polymer glutenins. In the semi-dilute regime, scattering experiments highlight two classes of behavior, akin to standard polymer solution and polymer gel, depending on the protein composition. We demonstrate that these two classes are encoded in the structural features of the proteins in very dilute solution, and are correlated with the presence of proteins assemblies of typical size tens of nanometers. The assemblies only exist when the protein mixture is sufficiently enriched in glutenins. They are found directly associated to the presence in the gel of domains enriched in non-exchangeable H-bonds and of size comparable to that of the protein assemblies. The domains are probed in neutron scattering experiments thanks to their unique contrast. We show that the sample visco-elasticity is also directly correlated to the quantity of domains enriched in H-bonds, showing the key role of H-bonds in ruling the visco-elasticity of polymer gluten gels.
\end{abstract}

%\maketitle

\section{Introduction}

Polymer materials may exhibit a large variety of unique properties, ranging from high water content, softness, and flexibility for hydrogels to resilience and temperature sensitivity for elastomers. The specific properties of polymer materials entail specific uses in different contexts including biomedical applications for hydrogels made of natural or biodegradable synthetic polymers~\cite{li_injectable_2012}, construction materials, and sensors for elastomers~\cite{Elastomers}.
Nature also abounds in polymer gels and elastomers with unique properties dedicated to specific functions, as mucus~\cite{meldrum_mucin_2018}, synovial fluid ~\cite{jay_role_2007}, the gelatinous layer of tension wood~\cite{nishikubo_xyloglucan_2007}, seed mucilage hydrogels~\cite{yu_multi-scale_2019} and natural rubber~\cite{Rubber}. In all these examples, the complex multi-component composition and the many types of interactions at play are intimately connected to drive the hierarchical structures of the polymer-like gel materials and control their mechanical properties.

This complexity and this intricacy also hold for gluten. Gluten, the insoluble protein of wheat, forms in its water hydrated state a highly cohesive and viscoelastic mass (comprising typically $2$ g of water per g of protein), akin to an elastomer~\cite{Gluten}. Gluten visco-elasticity is crucial in food science as it allows wheat flour to be baked into bread and biscuit. Gluten proteins belong to the broad family of prolamins; they are proteins rich in proline and glutamine amino-acids, which may confer texture to food materials because of the formation of protein strands under extensional flow~\cite{shewry_prolamin_1990}. Gluten is a complex mixture of several types of protein, which can be divided into two main classes, monomeric gliadins and polymeric glutenins. In glutenins, glutenin sub-units are linked together by disulfide bonds yielding polymers with molar mass up to several millions g/mol~\cite{FlourPower,wieser_chemistry_2007}. Gluten proteins belong to the wider class of intrinsically disordered proteins that are currently extensively investigated because of their crucial role in many biological processes. Gluten proteins are certainly the most documented
elastomeric plant proteins~\cite{tatham_wheat_2001}.  However, the study of gluten is very difficult because gluten proteins are broadly polydisperse and essentially insoluble in water. Hence, despite several decades of investigation, a full understanding of the structure of gluten in relation to its viscoelastic properties is still lacking. The active debate about gluten being regarded preferentially as a particulate gel or as a polymer gel is not fully closed (see~\cite{macritchie_theories_2014} and the references therein), although we strongly believe that the polymeric nature plays a major role, as inferred especially from our recent investigations~\cite{dahesh_polymeric_2014,dahesh_spontaneous_2016,banc_model_2017,morel_insight_2020,costanzo_tailoring_2020}. Previous works have also pointed out the important role of disulfide bonds and hydrogen bonds in the structuration of gluten~\cite{belton_mini_1999,morel_mechanism_2002,ng_large_2011}.

In order to shed light on the structure of gluten and on the relationship between structure and viscoelasticity, our strategy is to study model systems, produced by dispersing dedicated protein extracts in a good solvent, a mixture of water and ethanol, allowing an efficient solubilization of the proteins. Thanks to this approach, homogeneous samples with a wide range of protein concentration, spanning several orders of magnitude, can be studied. Note that this could not be performed with pure water as a solvent: water being a bad solvent for gluten, homogenous samples can only be produced at very high protein concentration. In our case, thanks to the choice of water/ethanol good solvent, unprecedented structural and mechanical data have been obtained. In particular, we have shown, for a given protein extract comprising equal amounts of gliadin and gliadin, that most structural and viscoelastic properties of protein dispersions can be qualitatively and quantitatively rationalized in the framework of polymer gels~\cite{dahesh_polymeric_2014, dahesh_spontaneous_2016}. More recently, we have developed a protocol to obtain from industrial gluten model protein extracts with controlled and tunable composition in gliadin and glutenin~\cite{MsExtraction}. In this paper, we leverage on this recent advancement and investigate the structure and visco-elasticity of suspensions of gluten proteins of various composition. This study allows us to demonstrate a correlation between the presence of large proteins assemblies in very dilute regime and the presence of domains enriched in H-bonds, of size comparable to the assemblies, in the semi-dilute regime, and a direct link between the hierarchical structures of proteins in the dilute regime and the sample visco-elasticity.

The paper is organized as follows. We first describe the materials and the different experimental methods. We then present and discuss the experimental results regarding the structure and visco-elasticity of the samples as probed thanks to a combination of complementary scattering and rheology techniques. We finally conclude by emphasizing the crucial role of protein assemblies in the materials.

\section{Materials and methods}

\subsection{Materials and sample preparation}
Gluten protein extracts are prepared from an industrial gluten (courtesy of TEREOS-SYRAL, Aalst, Belgium), following a protocol described elsewhere~\cite{morel_insight_2020,MsExtraction}. In brief, gluten is first dispersed at room temperature in a water/ethanol $50/50$ v/v solvent. Only the proteins well dispersed in the solvent are kept (the insoluble part is discarded). The dispersion is then quenched to a low temperature $T_q$, leading to a liquid-liquid phase separation into a light phase and a heavy phase. In the following, both the light and heavy phases are used after freeze-drying. The composition of the freeze-dried protein extracts are characterized by chromatography in a denaturating solvent, in which weak intra- and intermolecular interactions are suppressed~\cite{morel_insight_2020}. Overall, the light phase is enriched in gliadin, the monomeric proteins, whose molar mass, $M_{\rm{w}}$, lies in the range $(25-65) \times 10^3$ g/mol. Conversely, the heavy phase is enriched in glutenin, the polymeric proteins, with $M_{\rm{w}}$ from $90 \times 10^3$ g/mol to several $10^6$ g/mol. Interestingly, the exact composition of the protein extracts is varied by changing $T_q$, allowing a fine tuning of the protein composition.  In our experiments, $T_q$ ranges between $-0.8$ and $12.5^{\circ}$C. The composition of the protein extracts is characterized by its mass fraction of glutenin as
$GLU=\frac{m_{\rm{Glu}}}{m_{\rm{Glu}}+m_{\rm{Gli}}}$
with $m_{\rm{Glu}}$, respectively $m_{\rm{Gli}}$, the mass of glutenin, respectively gliadin, in the extract.  We obtain protein extracts with $GLU$ in the range $(1-66)$ \%.

Samples are prepared by dispersing the freeze-dried protein extracts in the appropriate volume of solvent, a water/ethanol $50/50$ v/v mixture. Hydrogenated and deuterated solvents are used. Deuterated solvent comprises OD ethanol (C$_2$H$_5$OD) and heavy water (D$_2$O). The protein concentration $C$ ranges between $4$ and $400$ mg/mL.

\subsection{Methods}

\subsubsection{Small-angle X-Ray scattering}
Small-angle X-ray scattering experiments are performed in house and in the European Radiation Synchrotron Facility, ESRF, (Grenoble, France).
The in-house set-up comprises a high brightness X-ray tube with low power and an aspheric multilayer optic (GeniX 3D from Xenocs) delivering an ultra low divergent beam ($0.5$ mrad); a two-dimensional Schneider 2D image plate detector prototype is used to collect the scattering intensity. The sample-detector distance is set at $1.9$ m. Synchrotron experiments are conducted at the ID02 beamline of ESRF~\cite{narayanan_multipurpose_2018}, using three different sample-detector distances ($d=1.5$, $7$ and $30$ m) in combination with a wavelength $0.0995$ nm, yielding scattering vectors $q$ in the range $(2 \times 10^{-3} - 7)$ nm$^{-1}$. In all experiments, the samples (prepared with hydrogenated or deuterated solvents and with protein concentration in the range ($10-400$ mg/mL) are held in glass capillaries of diameter $1.5$ mm. Standard non-linear fitting procedures are used to analyze the data.

\subsubsection{Small-angle and very small-angle neutron scattering}
Several facilities are used for small-angle neutron scattering experiments (SANS) and very small-angle neutron scattering experiments (VSANS).
SANS measurements at Laboratoire L\'eon Brillouin (Saclay, France) are performed on instrument PA20 using three configurations with the following wavelength $\lambda$ and sample-detector distance $d$ ($\lambda=0.6$ nm and $d=1.5$ m; $\lambda=0.6$ nm and $d=8$ m; $\lambda=1.5$ nm and $d=19$ m) yielding scattering vector $q$ in the range $(10^{-2} - 2)$ nm$^{-1}$.
VSANS and SANS are also conducted on two instruments operated by JCNS at the Heinz Maier-Leibnitz Zentrum (MLZ, Garching
Germany). SANS experiments are performed on KWS2~\cite{radulescu_studying_2016} using three configurations with $\lambda=0.7$ nm and $d=2$ m; $\lambda=0.7$ nm and $d=8$ m; $\lambda=1$ nm and $d=20$ m, yielding scattering vector $q$ in the range $(2 \times 10^{-2} - 3)$ nm$^{-1}$. VSANS experiments are performed at KWS3~\cite{heinz2015} with $\lambda=1.28$ nm and $d=10$ m yielding $q$ in the range ($2.2 \times 10^{-3} - 2 \times 10^{-2})$ nm$^{-1}$ and at KWS2  with $\lambda=0.7$ nm and $d=20$ m using the focusing mode with MgF2 lenses~\cite{Frielinghaus}, yielding $q$ in the range $(3 \times 10^{-3} - 3 \times 10^{-2})$ nm$^{-1}$.
In all cases, samples (prepared with hydrogenated or deuterated solvents and with protein concentration in the range ($210-302$ mg/mL)) are held in quartz Hellma cells with thickness of $1$ mm or $2$ mm. Standard data correction and calibration are  performed to analyze the data. Data are corrected for empty cell scattering, solvent scattering,  transmission, and detector sensitivity. Absolute scale transformation is performed using standard procedures~\cite{qtikws}. Standard non-linear fitting procedures are used to analyze the scattering profiles.

\subsubsection{Rheological measurements}

Linear viscoelastic measurements are performed using a Anton Paar MC302 stress-controlled rheometer. We use cone and plate geometries, with different diameters ($8$, $25$ or $50$ mm), depending on the sample visco-elasticity. The whole geometry is immersed in a bath of silicon oil to avoid solvent evaporation. After loading the sample (with a spatula for gels or simply by pouring liquid samples), the gap between the cone and the plate is set to its prescribed value ($100 \mu$m, for the $50$ mm diameter plate, $50 \mu$m for smaller plates). We let the samples equilibrate until the normal force acting on the cone relaxes to zero before starting the measurements. The frequency dependence storage, $G'$, and loss, $G"$, moduli are measured in the linear regime, with a typical strain amplitude of $1$ \%. Measurements are performed with samples prepared with deuterated solvents with a fixed protein concentration $C=237$ mg/mL and different $GLU$ content. The temperature is fixed at $25^{\circ}$C.

\section{Results and discussion}

\subsection{Structural features in very dilute regime}

\begin{figure}[h]
\centering
  \includegraphics[height=7 cm]{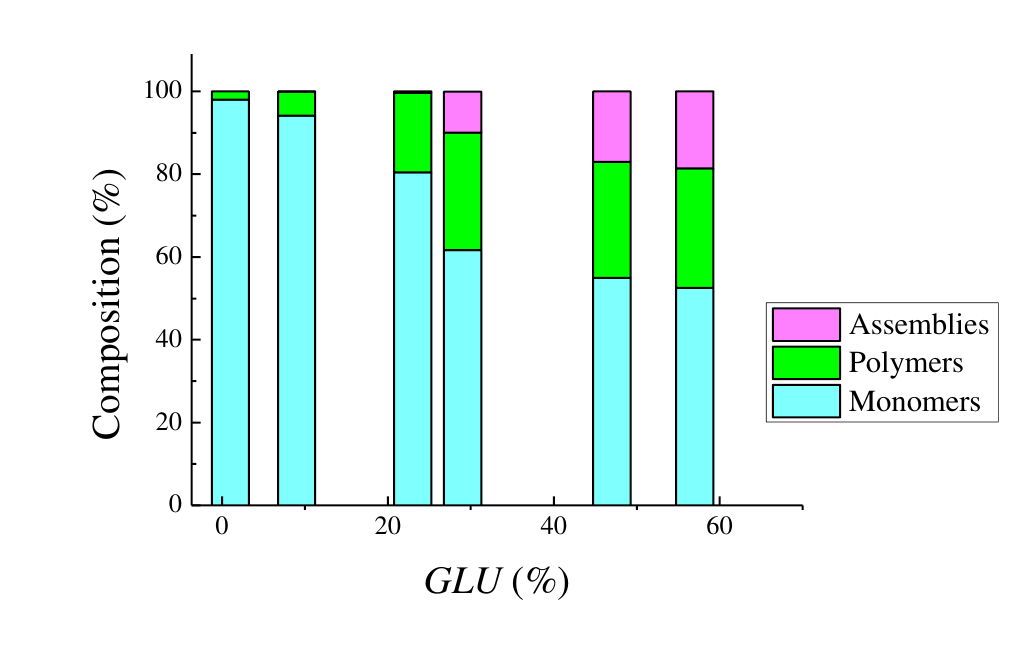}
 \caption{{Mass fraction of monomers, polymers and assemblies (see text) in dilute suspensions of the different protein extracts. Adapted from~\cite{morel_insight_2020}. The relative error, as evaluated from 3 replicated measurement with the extract with $GLU=47$ \% is less than $5$ \%. }}
 \label{fgr:A4F}
\end{figure}

Figure~\ref{fgr:A4F} reports the composition of the species present in very dilute suspensions ($C=4$ mg/mL) of the different protein extracts as determined thanks to asymmetrical
flow field-flow fractionation~\cite{morel_insight_2020}. Three classes of objects, monomeric gliadins, glutenin polymers and protein assemblies, are identified depending on their average size, $<R>$, and molar mass $<M_{\rm{w}}>$. For gliadins $<R> = 7$ nm, $<M_{\rm{w}}> = 8 \times 10^4$ g/mol, for glutenin polymers $<R> = 20$ nm and $<M_{\rm{w}}> = 4 \times 10^5$ g/mol and for protein assemblies $<R> = 85$ nm, $<M_{\rm{w}}> = 3 \times 10^7$ g/mol~\footnote{Note that for gliadins and polymers, $<R>$ refers to a mean hydrodynamic radius, whereas it refers to a mean radius of gyration for the assemblies.}. We find that the proportion of the different species evolves with the protein composition. As anticipated, at small $GLU$ ($GLU < $ ca $30$ \%), the proportion of polymers increases with $GLU$. Interestingly, above $30$ \%, the proportion of polymer is roughly constant and one observes the emergence of large protein assemblies as a third species, whose amount increases with $GLU$.

\subsection{Structural features in the semi-dilute regime}

\subsubsection{Spatial distribution of the gluten proteins}

\begin{figure}[h]
\centering
  \includegraphics{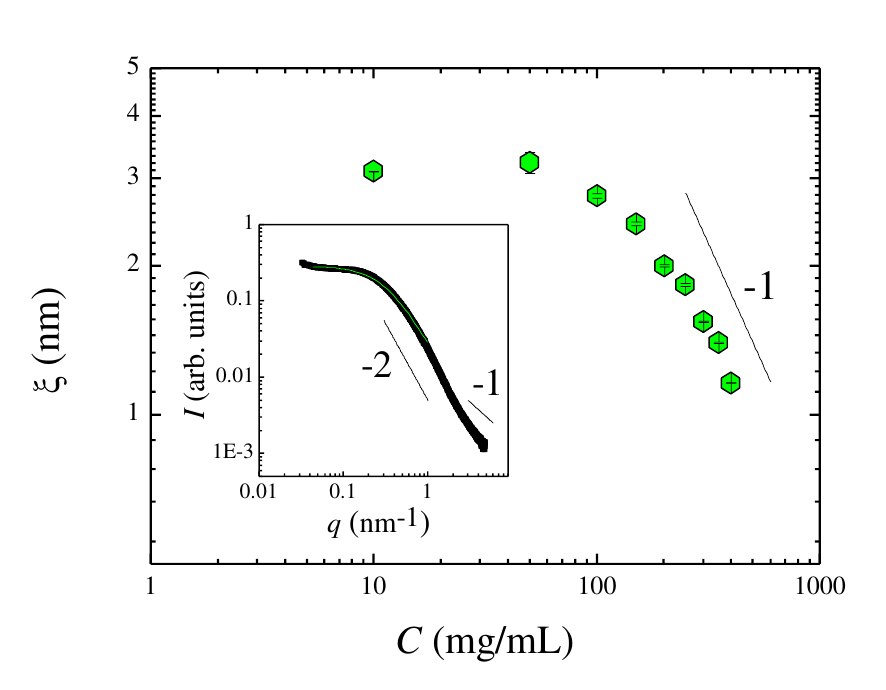}
 \caption{{Correlation length, $\xi$, as a function of the protein concentration for a protein extract with $GLU=13$ \%. $\xi$ is extracted from a fit with a Lorentzian function (see text). The inset shows the scattered intensity measured at room temperature as a function of the scattering vector for a sample with $C=10$ mg/mL and $GLU=13$ \%. Black symbols are experimental data points and the green solid line is the best fit. Measurements are performed at room temperature.}}
 \label{fgr:OZ}
\end{figure}

X-ray scattering is sensitive to the electronic densities of the species. Hence in small-angle X-ray scattering experiments, the contrast originates from the difference in the scattering densities between the gluten proteins and the solvent. Experiments therefore probes the spatial distribution of the proteins in the solvent.
We show (inset Fig.~\ref{fgr:OZ}) that the scattering profile of a low concentration sample depleted in glutenin ($C=10$ mg/mL, $GLU=13$ \%) is typical of a solution of polymer coils in the dilute regime. Note that this is in accordance with the fact that gluten proteins are intrinsically disordered proteins~\cite{hofmann_polymer_2012, kikhney_practical_2015, balu_effects_2016}. At intermediate scattering vectors ($0.5$ nm$^{-1}<q<3$ nm$^{-1}$), the scattered intensity, $I$, varies as $q^{-p}$, with $p=2$. This power law scaling is characteristic of Gaussian chains in a theta-solvent~\cite{ColbyRubinstein}. At smaller length scale, the transition from the $q^{-2}$ scaling to a $q^{-1}$ scaling, at a scattering vector $q_c$, allows the determination of the persistence length $l_p$ of the chains, following $q_c \times l_p =1.9$~\cite{denkinger_determination_1991}. We measure $q_c$ of the order of $2.8$ nm$^{-1}$ yielding $l_p$ of the order of $0.7$ nm. This small value indicates that the polypeptide chains of the gluten proteins are very flexible, which is typical for intrinsically disordered proteins~\cite{hofmann_polymer_2012,ohashi_experimental_2007}.
On the other hand, for length scales larger than the radius of gyration of the coils, at small $q$, a plateau of the scattered intensity is observed in a log-log representation.
By modeling the transition from the plateau to the power law decrease with a Lorentzian function as predicted with a Orstein-Zernicke (OZ) model, $I(q)=\frac{A}{1+(q\xi)^2}$, one can evaluate the correlation length $\xi$~\cite{ColbyRubinstein}. This length is equal to the radius of gyration of the scattering objects in dilute regime, and is expected to decrease with concentration in the semi-dilute regime. The fit with OZ (line in the inset Fig.~\ref{fgr:OZ}) gives $\xi=3.1$ nm, a numerical value consistent with the size of gliadin~\cite{morel_insight_2020,thomson_small_1999}. Note that data are equally well fitted using a Debye function~\cite{pedersen_scattering_2004}, which is the form
factor for Gaussian chains, yielding a comparable size ($4.6$ nm).

\begin{figure}[h]
\centering
  \includegraphics[height=7 cm]{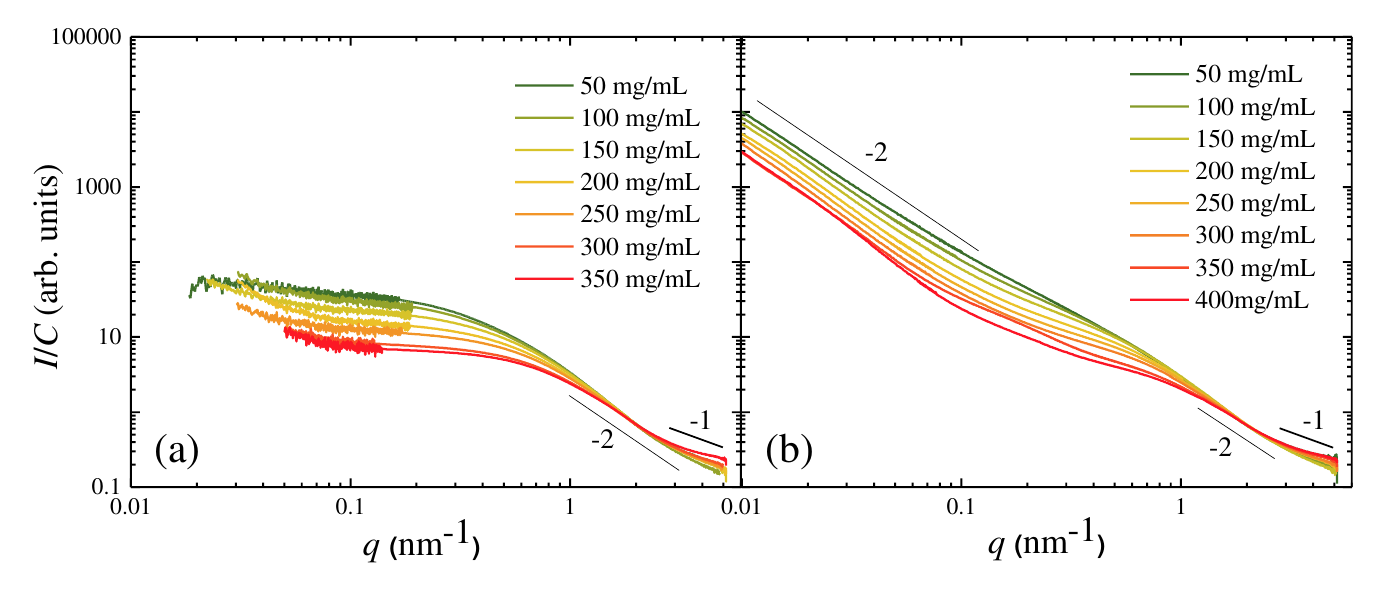}
 \caption{{Scattering profiles measured by small-angle X-ray scattering, for samples prepared with a hydrogenated solvent at various protein concentrations, as indicated in the legends, for protein extracts with (a) $GLU=13$ \%, and (b) $GLU=45$ \%. Measurements are performed at room temperature. Data are acquired at ESRF facility.}}
 \label{fgr:SAXS}
\end{figure}

Data acquired at different protein concentrations (Fig.~\ref{fgr:SAXS}a) all superpose at large scattering vectors, when the scattered intensity, $I$, is normalized by $C$, indicating a unique structure at small length scales, independent of the protein concentration, as expected. All data also exhibit a plateau at small $q$ but whose height, normalized by $C$ decreases as $C$ increases. This indicates a higher compressibility of the samples as $C$ increases, due to the interpenetration of the polymer coils, which signs the transition from dilute to semi-dilute regimes, at $C^*$. Above the overlap concentration $C^*$, the correlation length $\xi$ decreases due to coil interpenetration. We find that $\xi$, as obtained from fits of the scattering profiles with the OZ model, decreases from about  $3$ nm down to $1$ nm, as $C$ increases from $10$ to $400$ mg/mL (Fig.~\ref{fgr:OZ}). The overlap concentration $C^*$ is defined as the concentration from which $\xi$ decreases with increasing $C$. From purely geometrical arguments, the overlap concentration reads $C^*=\frac{3M_{\rm{w}}}{4 \pi R_{\rm{G}}^{3} Na}$, with $R_{\rm{G}}$ the radius of gyration of the coils, $M_{\rm{w}}$ their molar mass and $Na$ the Avogadro number. Experimentally, one measures $C^* \simeq 100 $ mg/mL. Taking $M_w=40000$ g/mol as average molar mass for gliadins, one determines a characteristic size of the order of $5.4$ nm, a numerical value in agreement with the values directly measured in the dilute regime. Finally, we note that the experimental evolution of $\xi$ with $C>C^*$ is consistent with the theoretically expected $C^{-1}$ scaling for a polymer in a theta-solvent~\cite{schaefer_unified_1984}, in full agreement with the $p=2$ power law exponent found for the scattered intensity at intermediate $q$-range.

Overall, glutenin depleted samples  exhibit the structural features of polymer coils in theta-solvent conditions. By contrast, the structural features of the samples enriched in glutenin polymer are more complex. We show in Figure~\ref{fgr:SAXS}b the scattering profiles, normalized by the protein concentration, for samples prepared with a gluten extract with $GLU=45$ \%. Data overlap at large $q$ and display the scattering features expected for polymer chains in theta-solvent, in agreement with what is measured for glutenin depleted samples (Fig.~\ref{fgr:SAXS}a). At larger length scale, i.e. at small $q$, on the other hand, the scattered intensity is found to vary as a power law with the scattering vector: $I \sim q^{-d_{\rm{f}}}$. This power law indicates large length scale heterogeneities in the spatial organization of the chains, which are characterized by a fractal dimension $d_{\rm{f}}$. The fractal dimension is the same at all concentrations but the amplitude of the power law slightly varies with $C$. The fractal structure is measured up to the smallest accessible $q$ ($q_{\rm{min}}=10^{-2}$ nm$^{-1}$), hence up to length scales of the order of $2\pi/q_{\rm{min}}\approx 600$ nm. This length scale is much larger than the typical size of the protein assemblies ($<R> = 85$ nm, see above). Considering the presence of much larger polymer-like objects for the samples prepared with $GLU=45$ \% than for the samples prepared with $GLU=13$ \%, we expect for the  samples with $GLU=45$ \% an overlap concentration smaller than the one for the samples with $GLU=13$ ($C^* \simeq 100 $ mg/mL). Hence, it is reasonable to state that data shown in Figure~\ref{fgr:SAXS}b very likely correspond to the semi-dilute regime.

\begin{figure}[h]
\centering
  \includegraphics[height=12 cm]{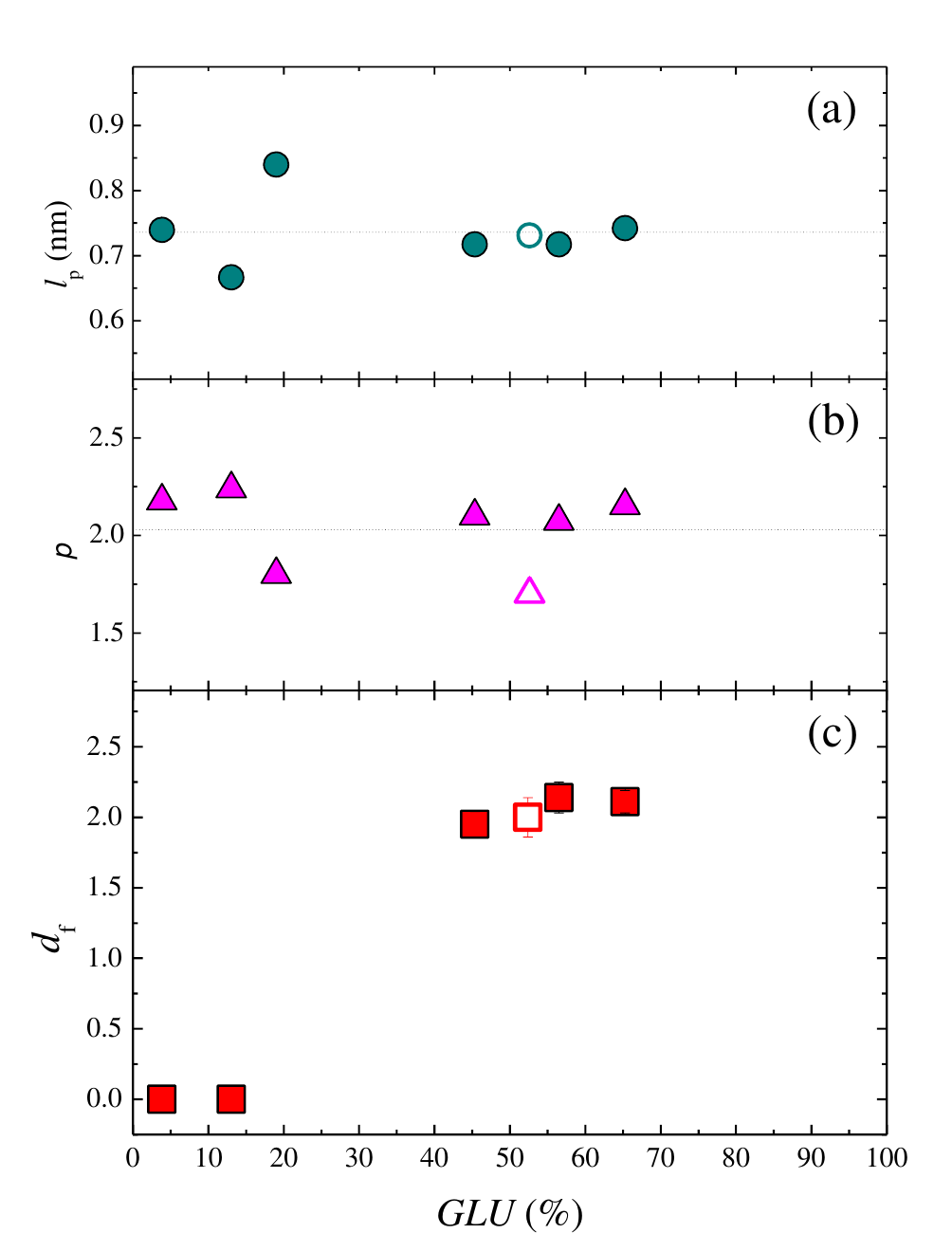}
 \caption{{Persistence length, $l_p$ (a), exponent of the scattered intensity with wave vector $q$ at large $q$, $p$ (b) and fractal dimension measured at small $q$ (c) as a function of the glutenin fraction of the protein extract. Open symbols correspond to data obtained with slightly different set-ups and protocols~\cite{dahesh_polymeric_2014}, which may explain the difference, in particular for the parameter $p$. The dotted lines in (a, b) correspond to the average of $l_p$ (a), and $p$ (b).}}
 \label{fgr:SAXSanalysis}
\end{figure}

Similar analysis as the ones described above are performed for samples prepared with different protein extracts, with $GLU$ ranging from $4$ to $66$ \%. The evolutions with the glutenin content of several structural parameters, the persistence length, the power law exponent at intermediate $q$, and the fractal dimension at smaller $q$, are plotted in Figure~\ref{fgr:SAXSanalysis}. Results show a polymer-like behavior of all extracts, independently of their composition, with a same persistence length ($l_p=0.74 \pm 0.1$ nm) and a same behavior at small length scale ($p=2.0 \pm 0.2$). Interestingly, however, two regimes are clearly evidenced, by the onset of a power law evolution of the scattered intensity at small $q$, characterized by a fractal dimension of the order of $2$. This onset coincides with the threshold for the presence of protein assemblies detected in very dilute regime ($GLU$ larger than $30$ \% typically).

\subsubsection{Indirect probing of H-bonds between proteins}

\begin{figure}[h]
\centering
  \includegraphics[height=7 cm]{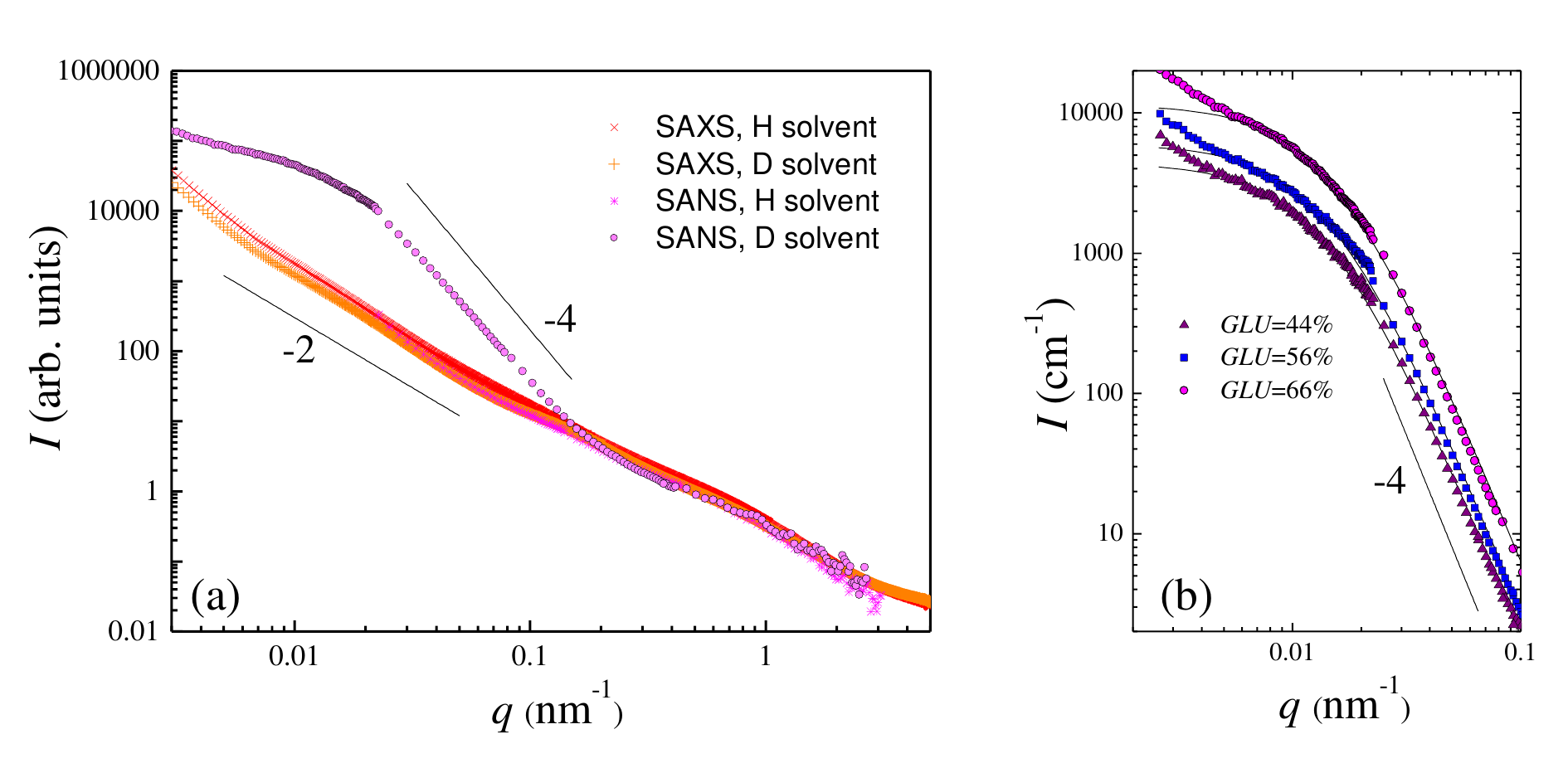}
 \caption{{Scattering profiles measured by (a) SAXS and VSANS/SANS, for a sample with $GLU=66$ \% prepared in a hydrogenated or deuterated solvent, (b) SANS and VSANS for samples prepared in deuterated solvent with different protein extracts as indicated in the legend. The symbols are data points and the lines are best fits using a Debye-Bueche model (see text). In (a, b) the protein concentration is $C=237$ mg/mL. In (a), because of different contrast in  SAXS and VSANS/SANS, data are shifted vertically to allow an overlap of the scattered intensity  at  large $q$. Measurements are performed at room temperature. SAXS data are acquired at ESRF facility and VSANS/SANS data are acquired at MLZ facility.}}
 \label{fgr:SAXSSANS}
\end{figure}

Figure~\ref{fgr:SAXSSANS}a summarizes the features of the scattering profiles, as measured by small-angle X-ray scattering (SAXS), and by very small-angle and small-angle neutron scattering (VSANS and SANS), for samples prepared in hydrogenated and deuterated solvents. Data are only displayed for fixed protein concentration ($C=237$ mg/mL) and composition ($GLU=66$ \%) but similar results are obtained for other $GLU$ (data not shown). As mentioned above, in SAXS, one probes the spatial distribution of the protein chains in the solvent. A neutron scattering experiment by contrast is sensitive to the scattering length densities of the different species, and the main contrast probed in the experiment changes depending whether the solvent is hydrogenated or deuterated. In the case of a hydrogenated solvent, the main contrast is the one between proteins and solvent. We find that the scattering profiles measured in SAXS (for both solvents) and in SANS using an hydrogenated solvent nicely overlap in the whole range of scattering vectors. The perfect overlap of the SAXS data for hydrogenated and deuterated solvents in the whole $q$-range demonstrates that replacing H by D in the solvent molecules does not change the spatial organization of the proteins, for length scales ranging from $\sim 1$ nm to $\sim 1$ $\mu$m, suggesting identical interactions at play. In the case of neutron scattering measurements with a deuterated solvent, the contrast mainly originates from the differences between the scattering length density of H and D. In that case, we observe that data at large $q$ ($q>0.2$ nm$^{-1}$) also overlap with the other scattering profiles, showing a unique structure of the protein chains at small length scale. In sharp contrast, however, drastically different scattering profiles are measured by VSANS and SANS for a deuterated solvent in the low $q$ region: instead of the $q^{-2}$ scaling, due to large scale fractal organization of heterogeneities in the proteins spatial organization, a $q^{-4}$ scaling followed by a pseudo-plateau at very small $q$ is measured. As already discussed previously for a sample with $GLU=52$ \%~\cite{banc_small_2016}, the striking difference between the scattering profiles originates from the heterogeneous exchange between the deuterium atoms comprised in the solvent and the labile hydrogen atoms of the protein chains. In certain regions of the sample, strong and/or multiple H-bonds between proteins prevent the standard D/H exchange between solvent and proteins. In these domains, one expects an enrichment in H as opposed to other parts of the samples. Neutron scattering experiments are sensitive to the H/D contrast, hence are probing the H-rich domains, which are domains enriched in H-bonds between proteins. The $q^{-4}$ scaling indicates well defined H-rich domains with sharp interfaces. The transition from this scaling towards a plateau at smaller $q$ allows an evaluation of the characteristic size of these domains. More quantitatively, the scattering profiles can be fitted with a Debye Bueche model (DB), conventionally used to describe micro-phase separated solids with sharp interfaces~\cite{debye_scattering_1949}: $I=\frac{I_0}{[1+(q\Xi)^2]^2}$. Here $I_0$ is the plateau value of the scattering intensity at low $q$ and $\Xi$ is the characteristic size of the phase-separated domains. We find that the DB model accounts well for the experimental data, as exemplified by the fits of some selected data (Fig.~\ref{fgr:SAXSSANS}b). We find moreover that
the characteristic size of the domains enriched in H-bonds between proteins are constant and independent on the protein composition, $\Xi=(67\pm2)$ nm. Interestingly, the size $\Xi$ is comparable to the size of the protein assemblies measured in very dilute solution (average size $85$ nm).

\begin{figure}[h]
\centering
  \includegraphics[height=7 cm]{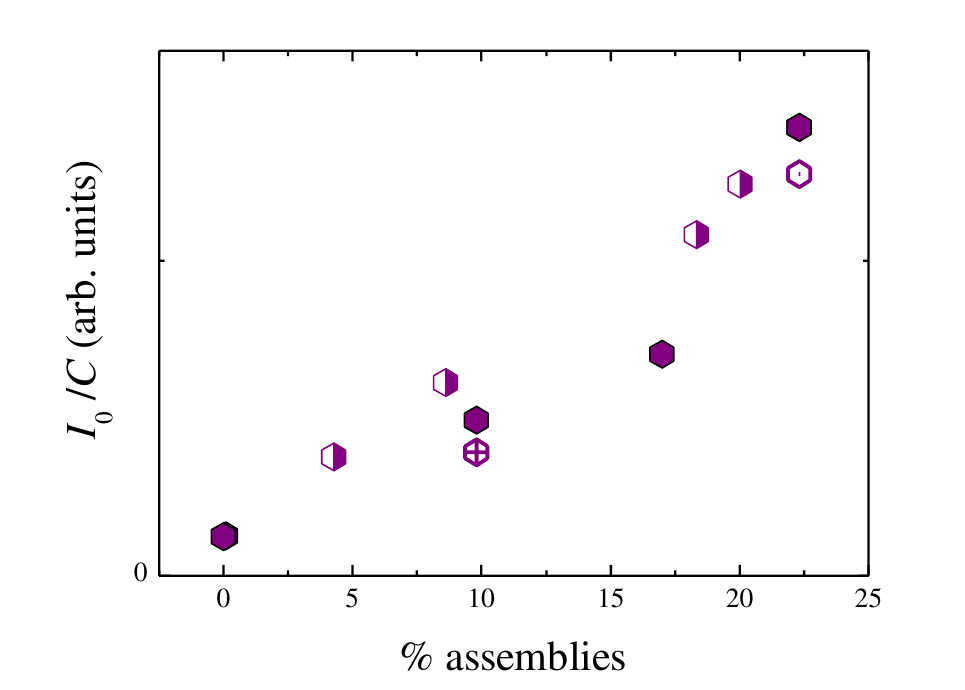}
 \caption{{Scattered intensity at very low $q$ as measured in VSANS normalized by the protein concentration as a function of the amount of assemblies in the samples. Full and empty symbols correspond to measurements performed with protein extracts as obtained with the protocol described in the text . Gluten concentration is $C=237$ mg/mL for full symbols (as in Fig.~\ref{fgr:SAXSSANS}b), for the cross-filled symbol (resp. dot-filled), $C=210$ mg/ml (resp. $302$ mg/ml) and $GLU=56$ \% (resp. $44$ \%). Half-filled symbols correspond to measurements performed with mixtures of two extracts with $GLU=13$ \% and $GLU=66$ \%.}}
 \label{fgr:Assemblies}
\end{figure}

To check more quantitatively the link between the domains enriched in H-bonds between proteins, measured in semi-dilute regime, and the presence of protein assemblies as inferred from measurements in very dilute regime, we plot in Figure~\ref{fgr:Assemblies} $I_0/C$, with $I_0$ the value of the low $q$ plateau of the scattered intensity, as a function of the \% of assemblies in the sample. The \% of assemblies is evaluated in dilute samples using asymmetrical
flow field-flow fractionation coupled to a differential
refractive index detection.
We measure that $I_0/C$ varies roughly linearly with the  \% of assemblies in the samples. Note that because the size of the domains is measured to be constant and independent of the composition of the protein extract, $I_0$ is expected to be directly proportional to the number of domains enriched H-bonds between proteins per unit volume in the sample (assuming a constant H/D contrast and a constant composition of the domains). Hence Figure~\ref{fgr:Assemblies} suggests a direct proportionality constant between the number of protein assemblies and the number of H-bonds rich domains. In Figure~\ref{fgr:Assemblies}, full symbols correspond to measurements performed with different protein extracts as obtained with the protocol described above with different quenching temperature $T_q$, and half-filled symbols correspond to VSANS measurements performed with mixtures of two extracts with $GLU=13$ \% and $GLU=66$ \%. The reasonable collapse of all data onto a single curve indicate that a simple dilution law holds. It moreover suggests that the protein assemblies are stable whatever their environment (quantity of solvent and presence of gliadins).

\begin{figure}[h]
\centering
  \includegraphics{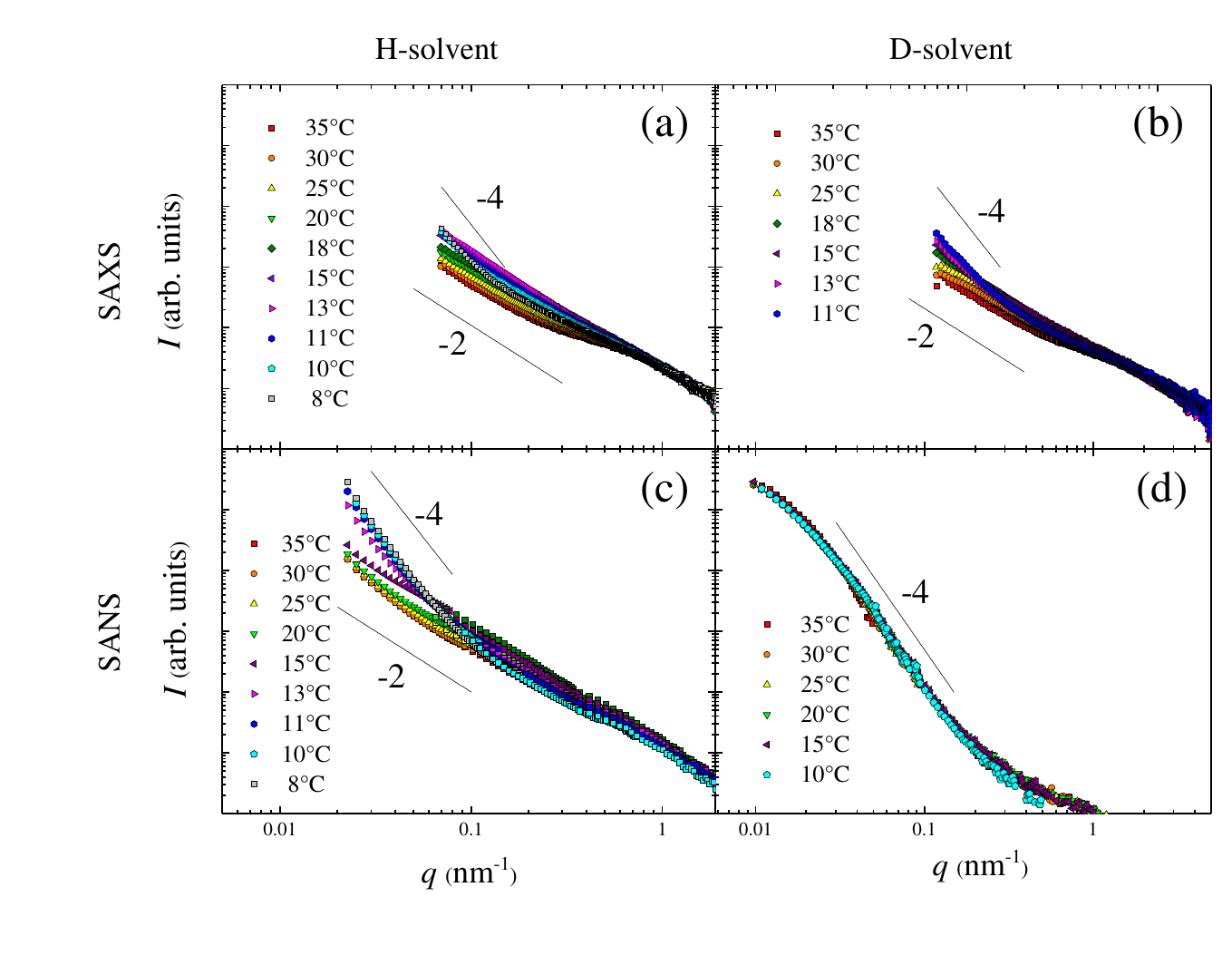}
 \caption{{Scattering profiles measured at different temperatures, as indicated in the legend, by small angle X-ray scattering (a, b) and neutron scattering (c, d), for samples with a fixed protein concentration ($C=237$ mg/mL) and a fixed glutenin content ($GLU=66$ \%) prepared in a hydrogenated solvent (a, c) or deuterated solvent (b, d). SAXS measurements have been acquired using a in-house set-up and SANS measurements have been acquired at LLB (c) and MLZ (d) facilities.}}
 \label{fgr:Temperature}
\end{figure}

To better assess the stability of the regions enriched in H-bonds between proteins, we investigate the evolution of the scattering profiles with temperature. Measurements acquired at different temperatures, from $8$ to $35^{\circ}$C, in SAXS and SANS and for hydrogenated and deuterated solvents, are shown in Figure~\ref{fgr:Temperature} for a sample with $C=237$ mg/mL and $GLU=66$ \%. As temperature decreases, the onset of a liquid-liquid phase-separation is evidenced in SAXS by a transition from a $q^{-2}$ to a $q^{-4}$ power law dependence at small $q$. As anticipated from previous experiments in a fully hydrogenated solvent~\cite{banc_phase_2019}, the liquid-liquid phase separation is detected, both by SAXS and SANS, for samples prepared with a hydrogenated solvent (H-solvent). We find that a liquid-liquid phase separation also takes place for a sample prepared in a deuterated solvent (D-solvent). For a sample prepared in a H-solvent, slightly different temperatures are determined with the two techniques (which might be explained by the different set-ups used). Using a same SAXS apparatus, we measure a significantly higher temperature in a D-solvent (about $18^{\circ}$C) than in a H-solvent (about $14^{\circ}$C), consistently with previous measurements by differential scanning calorimetry for other samples~\cite{costanzo_tailoring_2020}. This is in line with differences in the strength of hydrogen bonds for H and for D and hints as a role of hydrogen-bonding as driving force for liquid-liquid phase-separation. Interestingly, we do not find any modification with temperature of the SANS pattern of a sample prepared in a D-solvent (Fig.~\ref{fgr:Temperature}d). Hence, the number and size of the domains enriched in H-bonds between proteins are not affected by the liquid-liquid phase separation. We believe these domains get concentrated in the rich-phase, which is enriched in glutenins~\cite{MsExtraction}, in agreement also with infrared data showing that the heavy phase in enriched in H (and depleted in D) as compared to the light phase~\cite{banc_phase_2019}. In addition, the fact that these domains are not perturbed when the temperature varies in the one-phase region (in the range $[18-35]^{\circ}$C) strongly suggests that they are very stable and robust structures.

\subsection{Linear visco-elasticity}

\begin{figure}[h]
\centering
  \includegraphics[height=9 cm]{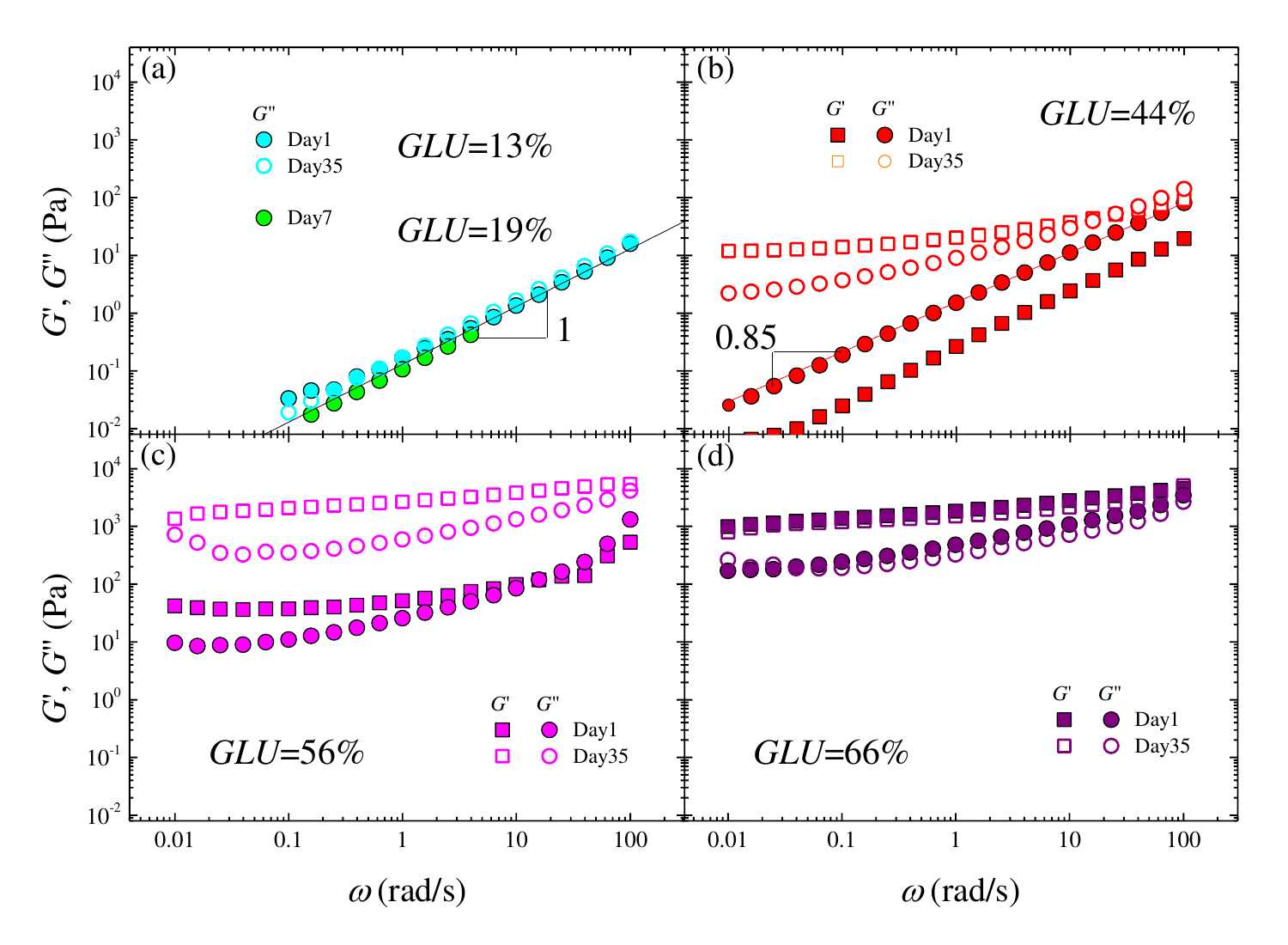}
 \caption{{Storage and loss moduli as a function of frequency, for fresh and $35$ days old samples prepared with a fixed protein concentration $C=237$ mg/mL, and with protein extracts comprising various amount of glutenin, as indicated in the legend. In (a, b), the lines are power law evolution with an exponent of $1$ (a) and $0.85$ (b).}}
 \label{fgr:RheoAging}
\end{figure}

We report in Figure~\ref{fgr:RheoAging}, the frequency dependence of the storage, $G'$, and loss, $G"$, moduli as a function of frequency, for samples with a fixed concentration ($C=237$ mg/mL), but different composition of the gluten extract.
Samples depleted in protein assemblies ($GLU=13$ \% and $GLU=19$ \%) are purely viscous. The storage modulus is too low to be measured reliably and the loss modulus is found to be proportional to the frequency: $G"= \eta \omega$, with a viscosity $\eta \simeq 100$ mPas. This value is in agreement with measurements performed with gliadin suspensions prepared using a different protocol~\cite{boire_osmotic_2015}.
By contrast, the samples prepared with protein extracts comprising protein assemblies ($GLU > 30$ \%) display a marked visco-elastic signature.
The two most enriched in glutenin samples ($GLU=56$ and $66$ \%) are essentially elastic: their storage modulus is nearly frequency-independent, and is larger than their loss modulus, in most or the whole experimentally investigated frequency range (from $10^{-2}$ to $10^{2}$ rad/s).

Concentration-dependent aging of gels prepared with $GLU=52$ \% has been previously investigated in detail by some of us and we have shown that the visco-elasticity and the gelation process could be quantitatively rationalized in the framework of near-critical gels~\cite{dahesh_spontaneous_2016,WinterChambon1987,martin_viscoelasticity_1988,martin_viscoelasticity_1989}. We show here that the same features occurs for a whole class of gluten gels. The near-critical gel features is especially exemplified in the sample with $GLU=44$ \%. In the window of experimentally accessible frequencies, a fresh sample exhibits the visco-elastic properties of a critical gel, with $G' \sim G" \sim \omega^{0.85}$ and $G" > G'$. For an aged gel on the other hand, an elastic plateau ($G'>G"$) is measured at low frequency and the transition to power evolution of the two moduli is measured at higher frequency. This is the signature of near-critical visco-elasticity above the gel point.

Visco-elasticity of the gluten gels is governed by H-bonds~\cite{dahesh_spontaneous_2016,ng_power_2007,costanzo_tailoring_2020}, and aging, as related to the increase of the elastic modulus with the time elapsed since sample preparation, is likely due to the reorganization of the H-bonds in the sample. In accordance, we note that the purely viscous samples do not exhibit any aging features, as opposed to the visco-elastic samples. Interestingly, we also find that the sample that is the most enriched in glutenin ($GLU=66$ \%) does not seem to exhibit any aging over the investigated period (data for a fresh sample and a $35$ day-old sample perfectly overlap), as opposed to the samples with $GLU=44$ \% and $GLU=56$ \%, which exhibit significant increase of their complex modulus with time. Although, this finding should deserve a deeper investigation, we believe this might be due to the hindrance of the H-bonds reorganization in a highly elastic material.

Overall we find that the rheological properties of the samples are directly related to their structure. The emergence of visco-elasticity is directly correlated to the presence of protein assemblies in the dilute regime, which is also associated to the presence of H-bonds-rich domains in semi-dilute regime.

\section{Conclusion}

We have investigated the structure of dispersion of gluten proteins, with tuneable composition, in a good solvent. Gluten proteins are mainly composed of a blend of monomeric proteins, gliadins, and polymeric proteins, glutenins. In principle, gluten proteins are by themselves at the cross-road between polymers and colloids. Despite the complexity and the numerous interactions at play in gluten, however, our experiments show that the material properties are dominated by the polymer nature of the constituents. Thanks to a combination of several techniques that probe the sample properties in different concentration regimes, we have evidenced the major role played by the protein assemblies. These assemblies are non compact and very stable structures with a size of the order of $100$ nm, which form even in very dilute regime, once the proportion of glutenins in the protein blend is sufficiently high. They can be assimilated to microgels with polymer chains held together by weak hydrogen bonds. Thanks to their distinctive contrast in neutron scattering, we have been able to identify their signature in semi-dilute regime and quantify their amount as a function of the initial protein composition. The fact that sizes of the same order of magnitude are measured when the protein concentrations varies by almost two orders of magnitude is intriguing and would deserve further investigation.  Finally, when varying the protein composition, we find that the emergence of visco-elasticity coincides with the emergence of the protein assemblies, demonstrating their crucial function tuning gluten gel mechanical properties.
Gluten being an essential ingredient of dough, and being the ingredient largely responsible for the unique visco-elastic properties of wheat dough, characterizing and rationalizing the properties of gluten gels is obviously crucial in many technological and industrial applications.

\section*{Acknowledgements}
Financial supports from ANR Elastobio (ANR 18 CE06 0012 01) and from Labex Numev
(ANR-10-LAB-20) are acknowledged.  This work is also based upon experiments performed at the KWS-2 and KWS-3
instruments operated by JCNS at the Heinz Maier-Leibnitz Zentrum
(MLZ), Garching, Germany, at PA-20 beamline  operated by Laboratoire Léon Brillouin, Gif-sur-Yvette, France and at ID02 beamline operated at the European Synchrotron Radiation Facility
(ESRF), Grenoble, France. The authors thank Theyencheri Narayanan and
Alessandro Mariani for assistance in using beamline ID02 at ESRF and
Philippe Dieudonné for help in the in-house SAXS measurements.

\section*{Bibliography}

\bibliography{gluten-JPCM}

\providecommand{\newblock}{}
\begin{thebibliography}{10}
\expandafter\ifx\csname url\endcsname\relax
  \def\url#1{{\tt #1}}\fi
\expandafter\ifx\csname urlprefix\endcsname\relax\def\urlprefix{URL }\fi
\providecommand{\eprint}[2][]{\url{#2}}

\bibitem{li_injectable_2012}
Li Y 2012 {\em Chemical Society Reviews\/} {\bf 41} 2193–2221

\bibitem{Elastomers}
Cankaya N 2017 {\em Elastomers\/} (IntechOpen)

\bibitem{meldrum_mucin_2018}
Meldrum O~W, Yakubov G~E, Bonilla M~R, Deshmukh O, McGuckin M~A and Gidley M~J
  2018 {\em Scientific Reports\/} {\bf 8} 5802

\bibitem{jay_role_2007}
Jay G~D, Torres J~R, Warman M~L, Laderer M~C and Breuer K~S 2007 {\em
  Proceedings of the National Academy of Sciences\/} {\bf 104} 6194--6199

\bibitem{nishikubo_xyloglucan_2007}
Nishikubo N, Awano T, Banasiak A, Bourquin V, Ibatullin F, Funada R, Brumer H,
  Teeri T~T, Hayashi T, Sundberg B and Mellerowicz E~J 2007 {\em Plant and Cell
  Physiology\/} {\bf 48} 843--855

\bibitem{yu_multi-scale_2019}
Yu L, Yakubov G~E, Gilbert E~P, Sewell K, van~de Meene A~M and Stokes J~R 2019
  {\em Carbohydrate Polymers\/} {\bf 207} 333--342

\bibitem{Rubber}
Sabu T, Chin H~C, Laly P, Rajisha K~R, Jithin J and Maria H 2013 {\em Natural
  Rubber Materials\/} (RSC Polymer Chemistry Series)

\bibitem{Gluten}
Wrigley C, Bekes F and Bushuk W 2006 {\em Gliadin and Glutenin: The Unique
  Balance of Wheat Quality\/} (American Association of Cereal Chemists, Inc.)

\bibitem{shewry_prolamin_1990}
Shewry P~R and Tatham A~S 1990 {\em Biochemical Journal\/} {\bf 267} 1--12

\bibitem{FlourPower}
Wrigley C~W 1996 {\em Nature\/} {\bf 381} 738--739

\bibitem{wieser_chemistry_2007}
Wieser H 2007 {\em Food Microbiology\/} {\bf 24} 115--119

\bibitem{tatham_wheat_2001}
Tatham A~S, Hayes L, Shewry P~R and Urry D~W 2001 {\em Biochimica et Biophysica
  Acta (BBA)-Protein Structure and Molecular Enzymology\/} {\bf 1548} 187--193

\bibitem{macritchie_theories_2014}
MacRitchie F 2014 {\em Journal of Cereal Science\/} {\bf 60} 4--6 ISSN 07335210

\bibitem{dahesh_polymeric_2014}
Dahesh M, Banc A, Duri A, Morel M~H and Ramos L 2014 {\em J. Phys. Chem. B\/}
  {\bf 118} 11065--11076

\bibitem{dahesh_spontaneous_2016}
Dahesh M, Banc A, Duri A, Morel M~H and Ramos L 2016 {\em Food Hydrocolloids\/}
  {\bf 52} 1--10

\bibitem{banc_model_2017}
Banc A, Dahesh M, Wolf M, Morel M~H and Ramos L 2017 {\em Journal of Cereal
  Science\/} {\bf 75} 175--178

\bibitem{morel_insight_2020}
Morel M~H, Pincemaille J, Chauveau E, Louhichi A, Violleau F, Menut P, Ramos L
  and Banc A 2020 {\em Food Hydrocolloids\/} {\bf 103} 105676

\bibitem{costanzo_tailoring_2020}
Costanzo S, Banc A, Louhichi A, Chauveau E, Wu B, Morel M~H and Ramos L 2020
  {\em Macromolecules\/} {\bf 53} 9470--9479

\bibitem{belton_mini_1999}
Belton P~S 1999 {\em Journal of Cereal Science\/} {\bf 29} 103--107

\bibitem{morel_mechanism_2002}
Morel M~H, Redl A and Guilbert S 2002 {\em Biomacromolecules\/} {\bf 3}
  488--497

\bibitem{ng_large_2011}
Ng T~S~K, McKinley G~H and Ewoldt R~H 2011 {\em Journal of Rheology\/} {\bf 55}
  627--654

\bibitem{MsExtraction}
Pincemaille J, Lecacheaux L, Banc A, Menut P, Ramos L and Morel M~H {\em
  manuscript in preparation\/}

\bibitem{narayanan_multipurpose_2018}
Narayanan T, Sztucki M, Van~Vaerenbergh P, Léonardon J, Gorini J, Claustre L,
  Sever F, Morse J and Boesecke P 2018 {\em Journal of Applied
  Crystallography\/} {\bf 51} 1511--1524

\bibitem{radulescu_studying_2016}
Radulescu A, Szekely N~K, Appavou M~S, Pipich V, Kohnke T, Ossovyi V, Staringer
  S, Schneider G~J, Amann M, Zhang-Haagen B, Brandl G, Drochner M, Engels R,
  Hanslik R and Kemmerling G 2016 {\em Journal of Visualized Experiments\/}
  54639

\bibitem{heinz2015}
{Pipich, V and Fu, Z} 2015 {\em Journal of Large-Scale Research Facilities\/}
  {\bf 1} A31

\bibitem{Frielinghaus}
Frielinghaus H, Pipich V, Radulescu A, Heiderich M, Hanslik R, Dahlhoff K,
  Iwase H, Koizumi S and Schwahn D 2009 {\em Journal of Applied
  Crystallography\/}  681--690

\bibitem{qtikws}
\url{http://qtisas.com/}

\bibitem{hofmann_polymer_2012}
Hofmann H, Soranno A, Borgia A, Gast K, Nettels D and Schuler B 2012 {\em
  Proceedings of the National Academy of Sciences\/} {\bf 109} 16155--16160

\bibitem{kikhney_practical_2015}
Kikhney A~G and Svergun D~I 2015 {\em FEBS Letters\/} {\bf 589} 2570--2577

\bibitem{balu_effects_2016}
Balu R, Mata J~P, Knott R, Elvin C~M, Hill A~J, Choudhury N~R and Dutta N~K
  2016 {\em The Journal of Physical Chemistry B\/} {\bf 120} 6490--6503

\bibitem{ColbyRubinstein}
Rubinstein M and Colby R~H 2003 {\em Polymer Physics\/} (Oxford University
  Press, Oxford)

\bibitem{denkinger_determination_1991}
Denkinger P and Burchard W 1991 {\em Journal of Polymer Science Part B: Polymer
  Physics\/} {\bf 29} 589--600

\bibitem{ohashi_experimental_2007}
Ohashi T, Galiacy S~D, Briscoe G and Erickson H~P 2007 {\em Protein Science\/}
  {\bf 16} 1429--1438

\bibitem{thomson_small_1999}
Thomson N~H, Miles M~J, Popineau Y, Harries J, Shewry P and Tatham A~S 1999
  {\em Biochimica et Biophysica Acta (BBA)-Protein Structure and Molecular
  Enzymology\/} {\bf 1430} 359--366

\bibitem{pedersen_scattering_2004}
Pedersen J~S and Schurtenberger P 2004 {\em Journal of Polymer Science Part B:
  Polymer Physics\/} {\bf 42} 3081--3094 ISSN 0887-6266, 1099-0488
  \urlprefix\url{http://doi.wiley.com/10.1002/polb.20173}

\bibitem{schaefer_unified_1984}
Schaefer D~W 1984 {\em Polymer\/} {\bf 25} 387--394

\bibitem{banc_small_2016}
Banc A, Charbonneau C, Dahesh M, Appavou M~S, Fu Z, Morel M~H and Ramos L 2016
  {\em Soft Matter\/} {\bf 12} 5340--5352

\bibitem{debye_scattering_1949}
Debye P and Bueche A~M 1949 {\em Journal of Applied Physics\/} {\bf 20}
  518--525

\bibitem{banc_phase_2019}
Banc A, Pincemaille J, Costanzo S, Chauveau E, Appavou M~S, Morel M~H, Menut P
  and Ramos L 2019 {\em Soft Matter\/} {\bf 15} 6160--6170

\bibitem{boire_osmotic_2015}
Boire A, Menut P, Morel M~H and Sanchez C 2015 {\em The Journal of Physical
  Chemistry B\/} {\bf 119} 5412--5421

\bibitem{WinterChambon1987}
Winter H~H and Chambon F 1987 {\em Journal of Rheology\/} {\bf 30} 367--382

\bibitem{martin_viscoelasticity_1988}
Martin J~E, Adolf D and Wilcoxon J~P 1988 {\em Physical Review Letters\/} {\bf
  61} 2620

\bibitem{martin_viscoelasticity_1989}
Martin J~E, Adolf D and Wilcoxon J~P 1989 {\em Physical Review A\/} {\bf 39}
  1325--1332

\bibitem{ng_power_2007}
Ng T~S~K and McKinley G~H 2008 {\em Journal of Rheology\/} {\bf 52} 417--449

\end{thebibliography}
\bibliographystyle{iopart-num}

%\begin{thebibliography}{#1}"&Start of numeric reference list\\
%\end{thebibliography}"&End of numeric reference list\\

\end{document}